\documentclass[preprint2]{aa} 
\usepackage{graphicx}
\usepackage{cuted}
\usepackage{txfonts}

\usepackage{hyperref}                                   

\usepackage{natbib}
\usepackage[export]{adjustbox}

\bibpunct{(}{)}{;}{a}{}{,} 
\newcommand{\1}{{~\sc i}}
\newcommand{\2}{{~\sc ii}}
\newcommand{\3}{{~\sc iii}}
\newcommand{\4}{{~\sc iv}}
\newcommand{\5}{{~\sc v}}
\newcommand{\6}{{~\sc vi}}

\newcommand{\kms}{{\,km\,s$^{-1}$}}
\newcommand{\cc}{{\,cm$^{-3}$}} 
\newcommand{\mic}{{\,$\mu$m}}
\newcommand{\hmol}{H$_2$}

\newcommand{\eprint}[1][]{#1}

\usepackage[normalem]{ulem}
\usepackage{tikz}
\usepackage[export]{adjustbox}

\begin{document}

 \title{ESA Voyage 2050 White Paper\\A complete census of the gas phases in and around galaxies: far-UV spectropolarimetry as a prime tool for understanding galaxy evolution and star formation}

   
	   \titlerunning{A complete census of the gas phases in and around galaxies}
   	\authorrunning{Lebouteiller et al.}

   \author{V. Lebouteiller\inst{1}, C. Gry\inst{2}, H. Yan\inst{3,4}, P. Richter\inst{5}, B. Godard\inst{6}, E. B. Jenkins\inst{7}, D. Welty\inst{8}, N. Lehner\inst{9}, P. Guillard\inst{10}, J. Roman-Duval\inst{8}, E. Roueff\inst{11}, F. Leone\inst{12,13}, D. Kunth\inst{10}, J. C. Howk\inst{9}, P. Boiss{\'e}\inst{10}, F. Boulanger\inst{14}, E. Bron\inst{11}, B. James\inst{8}, J. Le Bourlot\inst{11}, F. Le Petit\inst{11}, M. Pieri\inst{2}, V. Valdivia\inst{1} }
   \institute{$^1$ AIM, CEA, CNRS, Universit\'e Paris-Saclay, Universit\'e Paris Diderot, Sorbonne Paris Cit\'e, F-91191 Gif-sur-Yvette, France \email{vianney.lebouteiller@cea.fr}\\
     $^2$ Aix Marseille Univ., CNRS, CNES, LAM, Marseille, France \\
     $^3$ Deutsches Elektronen Synchrotron  (DESY), Platane-nallee 6, D-15738 Zeuthen, Germany \\
     $^4$ Institut f\"ur Physik und Astronomie, Universit\"at Potsdam, D-14476 Potsdam, Germany \\
     $^5$ Institut f\"ur Physik und Astronomie, Universit\"at Potsdam, Karl-Liebknecht-Str. 24/25, D-14476, Golm, Germany \\
     $^6$ Observatoire de Paris, {\'E}cole normale sup{\'e}rieure, Universit{\'e} PSL, Sorbonne Universit{\'e}, CNRS, LERMA, F-75005, Paris, France \\
     $^7$ Dept. of Astrophysical Sciences, Princeton University, Princeton, NJ, USA \\
     $^8$ Space Telescope Science Institute, 3700 San Martin Dr, Baltimore, MD 21218, USA \\
     $^9$ Department of Physics, University of Notre Dame, Notre Dame, IN 46556, USA \\
     $^{10}$ Sorbonne Universit{\'e}, CNRS, UMR 7095, Institut d'Astrophysique de Paris, 98 bis Bd Arago, 75014 Paris, France \\
     $^{11}$  LERMA,  Observatoire  de  Paris,  PSL  Research  University,  CNRS,  Sorbonne  Universit{\'e}s,  UPMC  Univ.  Paris  06, F-92190 Meudon, France  \\
     $^{12}$ INAF-Osservatorio Astrofisico di Catania, Via S. Sofia 78, I-95123, Catania, Italy \\
     $^{13}$ Universit\'a degli studi di Catania, Via S. Sofia 78, I-95123, Catania, Italy \\
     $^{14}$ LERMA/LRA, Observatoire de Paris, PSL Research University, CNRS, Sorbonne Universit{\'e}, UPMCUniversit{\'e} Paris 06, Ecole Normale Sup{\'e}rieure, 75005 Paris, France \\
   }

   \date{White paper submitted: Aug. 6th, 2019.
   \vspace{-1cm} }


   \maketitle

  \begin{strip}
            \par 
        
  \section{Abstract}
  The wavelength range $912-2000$\,\AA\ (hereafter far-UV) provides access to atomic and molecular transitions of many species the interstellar medium (ISM), circumgalactic medium (CGM), and intergalactic medium (IGM), within phases spanning a wide range of ionization, density, temperature, and molecular gas fraction.
    
Far-UV space telescopes have enabled detailed studies of the ISM in the Milky Way, paving the way to understand, in particular, depletion of elements by condensation onto dust grains, molecular gas formation pathways, ionization processes, and the dynamics of shocks. Absorption features appearing in the UV spectra of hot stars yield fundamental insights into the composition and physical characteristics of all phases of the ISM along with the processes that influence them. \textit{However, no single instrument has as yet given access to species in all ISM phases at the same high spectral resolution}: from the molecular bands of CO and \hmol, to the cold neutral medium tracers (e.g., C\1, S\1), the warm medium tracers (e.g., C\2, N\1, O\1, Mg\2, Fe\2, Si\2\ etc...), and to the multiply charged ions of the hot ionized medium (e.g., C\4, Si\4, as well as O\6).

A comprehensive study of ISM phases, their interaction, and the nature of their boundaries requires comparing abundances and velocity profiles of tracers within these different phases but we have yet to design the spectrometer able to observe the full UV domain at resolving power $R>100\,000$ and detectors that can reach a signal-to-noise ratio SNR$>500$. The line FWHM being governed by turbulence, temperature, and species mass, such a resolution is necessary to resolve lines from both the cold molecular hydrogen and the warm metal ions with a turbulent velocity of $\approx1$ \kms, and to differentiate distinct velocity components, typically separated by less than $2$\,\kms. Future UV spectroscopic studies of the Milky Way ISM must \textit{revolutionize our understanding of the ISM as a dynamical, unstable, and magnetized medium, and rise to the challenge brought forward by current theories}. In particular, can we obtain observational signatures of the dissipation of turbulence in a magnetized medium, and how does the magnetic field help in structuring the ISM and in the transport of matter between phases? At stake is the full understanding of the atomic-to-molecular transition, the molecular complexity, and the various associated diagnostics that are currently used from dissipative to galactic scales.\\
\textit{[Cont'd next page.]}

Another interesting prospect is to \textit{transpose the same level of details that has been reached for the Milky Way to the ISM in external galaxies}, in particular in metal-poor galaxies, where the ISM chemical composition, physical conditions, and topology (arrangement of various phases) change dramatically, with significant consequences on the star-formation properties and on the overall galaxy evolution. We need to know, in particular, how star formation proceeds in quasi-pristine or pristine environments and what is the role of accreting clouds and compact objects in regulating the star formation process. To circumvent systematic biases in column density determinations and to examine the ISM enrichment as a function of the environment, next far-UV missions should be versatile enough to observe stellar clusters and individual O/B stars at distances of few Mpc to few $100$s\,Mpc respectively, with a spectral resolution power $R\sim10^5$ and $10^4$ respectively.

Such requirements are also necessary to perform statistical analyses of background quasar lines of sight intersecting the CGM of galaxies at various redshifts. The CGM is an important component of galaxies that connects the galaxy to the cosmic web and is, as such, at the center of many processes that we do not yet fully understand. With future UV missions, we ought to be able to  \textit{fill the gap between the various physical scales and phases and to comprehend the role of gas exchanges and flows for galaxy evolution}.  

We advocate a far-UV space telescope that will be able to tackle these issues. Such an observatory would enable access to the range $\approx900-3100$\,\AA, and would require an exceptionally large mirror $>5$\,m. The optimal observation mode would be spectropolarimetry with a resolution of $R\gtrsim10^5$.

\bigskip

\textbf{This ESA white paper draws from and expands on the following white papers previously submitted for the Astro 2020 Decadal Survey, \cite{Gry2019a}, \cite{Lebouteiller2019a}, and \cite{Yan2019a}. The original ESA white paper was reformatted as a 2-column paper for the present document.}

            \par
            \hfill
            \end{strip}

\section{Understanding the chemistry and physics of the ISM: The Milky Way laboratory}

\textbf{
\begin{itemize}
\item \textit{How is the ISM structured? Is the structure of molecular clouds compatible with our current understanding of physical and chemical processes in the ISM?}
\item \textit{What are the gas ionizing and heating mechanisms and what consequences for the chemistry and for star formation?}
\item \textit{How do the cold and hot medium trade matter and entropy at their interfaces?}
  \end{itemize}}

\bigskip

The Milky Way ISM provides the prime reference to understand the basic processes occuring in the ISM. While it is important to consider and to examine the ISM as an astrophysical object by itself, a profound understanding of its properties is also expected to lead to parameters that are relevant for star formation and galaxy evolution.

Matter in the interstellar space has long been considered to be distributed in diverse, but well defined phases that consist of (1) the hot, ionized interstellar medium (HIM; $T\sim10^{6-7}$\,K) emitting soft X-rays, (2) the warm-hot ionized medium (WHIM; $\sim10^{4-6}$\,K), (3) the warm neutral or ionized medium (WNM+WIM) ($6000-10^4$\,K), and (4) the cold ($\sim10-200$\,K) neutral medium (CNM) and molecular star-forming clouds that occupy only a few percent of the volume (e.g., \citealt{Field1969a,Bialy2019a}). The different phases are supposed to be globally consistent with hydrostatic equilibrium \citep{Ferriere1998a}, but the multi-phase aspect of the ISM is not fully understood and
non-equilibrium conditions are often contemplated. Observations do seem to point out that non-equilibrium conditions might be key to describe the observed abundances. How the different phases are related, how mass flows from one phase to the other, and how the transition from H$^0$ to \hmol\ occurs -- which is the first step in the path way leading to the formation of more complex molecules -- are still open questions. 

A comprehensive study of ISM phases and the nature of their boundaries or connections requires comparing abundances and velocity profiles from tracers of the different phases. In the UV, studies of a wealth of absorption features appearing in the spectra of hot stars yield fundamental insights into the composition and physical characteristics of all phases of the ISM along with the processes that influence them. They also inform us on the nature of boundaries between them. However no single UV instrument has as yet given access to species in all ISM phases at the same high spectral resolution: from the molecular bands of CO and H$_2$ to the highly ionized transitions that trace the HIM (C\4, Si\4, as well as O\6), and passing through the CNM (traced by C\1\ and S\1) and the WIM tracers (like C\2, N\1, O\1, Mg\2, Fe\2, Si\3...). 
The line width being governed by turbulence, temperature, and mass, a resolution power $R=\Delta\lambda/\lambda>120\,000$ is necessary to resolve cold atomic and molecular gas with a turbulent velocity of $1-2$\kms. \textit{We have yet to design the instrument that observes at $R>120\,000$ the full UV domain to study the multi-phase ISM.}

\subsection{Cold, dense and molecular gas}

The H$_2$ molecule is extremely difficult to detect and until now only few missions have been able to trace it directly (in particular \textit{Copernicus} and FUSE). H$_2$ can only be observed in two ways, (1) rovibrational transitions in the IR and (2) in absorption in the far-UV. Observing lines in absorption is the only way to measure atoms and molecules in their ground state, which is particularly important for \hmol, found essentially in the rotational excitation level $J=0$ and $1$ levels in the ISM. Infrared vibrational emission lines concern a negligible fraction of the total \hmol. Therefore a direct estimate of the mass of the cold molecular gas can only be obtained by observing \hmol\ in absorption in the far UV. Such a direct estimate is necessary in order to confirm and improve the various recipes (e.g., based on CO or dust measurements) used to infer the mass of H$_2$ and in turn to relate to the star formation process (e.g., \citealt{Narayanan2012a}). At the same time, a detailed knowledge of the molecular gas physical conditions and chemical properties is necessary in order to understand the role of the environment (radiation, shocks...) for the star formation process. 

Basic processes of interest in the molecular ISM include: (1) H$^0$-to-\hmol\ transition that can be probed with a large sample of various low- to high-extinction targets and the role of internal spatial structures for the self-shielding process of diffuse \hmol, (2) chemistry in truly molecular regions (molecular fraction $f_{\rm H{_2}}\approx 1$, shielded from UV radiation by dust and \hmol) where ionization by penetrating cosmic rays plays an important role, (3) change of molecular composition with $A_V$ by reaching opacities where HCN and HNC exist, (4) variation of the CO/\hmol\ abundance ratio with cloud depth, (5) characterization of the C$^0$-to-CO transition (occurring around $A_V=2$ depending on the interstellar radiation field), and (6) internal velocity structure and measure of the turbulence in molecular clouds.

At the same time, observations of H$_2$ can be interpreted correctly only if the dynamical influence of the environment as well as the interplay between the thermal processes related to the formation and destruction of H$_2$ are accounted for \citep{Valdivia2016a,Valdivia2017a}. A significant fraction of warm H$_2$, heated by the local dissipation of turbulent kinetic energy, exists in the low-density gas, thereby reflecting the complex intermix between the warm and cold gas in molecular clouds. The warm out-of-equilibrium H$_2$ is especially important for the formation of molecular species with high endothermicity, such as CH$^+$ \citep{Nehme2008a}. Section\,\ref{sec:dynamic} specifically discusses the dynamical aspects in the diffuse ISM. 

Carbon chemistry, in particular the abundance of C$^0$ can also be examined by looking for an expected discontinuity in obscured spectra at $1102$\,\AA\ due to the carbon photoionization continuum \citep{Rollins2012a}. For $A_V=2$ (resp.\ $A_V=3$) a flux decrease by a factor of $10$ (resp.\ $100$) is expected. Such a discontinuity depends on the relative abundance of C$^0$, governed by $n_{\rm H}$, $A_V$, and chemistry. Reaching higher $A_V$ ($\gtrsim2$) than those probed with FUSE is therefore required. It must also be noted that carbon ionization also depends on the abundance and charge of polycyclic aromatic hydrocarbons \citep{Kaufman2006a}. Column densities and abundance ratios of C$^0$, C$^+$, CO, and other molecules such as OH, H$_2$O, and C$_2$ (with far-UV transitions) or CH and CH$^+$ (in the visible) will then have to be compared to detailed photo-chemical models (e.g., Meudon PDR; \citealt{LePetit2006a}), dynamical models (e.g., Paris-Durham Shock model, \citealt{Lesaffre2013a,Godard2019a}), and numerical simulations of multiphasic turbulent ISM (e.g., \citealt{Valdivia2016a}).

High resolution of the order of $1-2$\,\kms\ is required to disentangle individual velocity components and separate fine-structure lines or H$_2$ rotational and rovibrational levels. \hmol, in particular, has been observed at a resolution better than $10$\,\kms\ only in a few lightly-reddened -- thus low $N({\rm H}_2)$ -- Galactic lines of sight \citep[with IMAPS, e.g.,][]{Jenkins1997a}.  Paradoxically, higher spectral resolution has been more readily available in the distant universe, due to redshifts that bring \hmol\ lines into the visible spectrum \citep[e.g.,][]{Noterdaeme2007a}. High resolution in the far-UV is absolutely necessary to study the physical conditions, the excitation and formation mechanisms of \hmol\ in the local universe. In more reddened sight lines it will be challenging to distinguish individual components in the $J=0,1$ lines which will be damped, but they could be resolved in the higher-$J$ level lines.
  
High sensitivity is also required to reach  high-extinction targets since we need to measure abundances in clouds with various opacities, and access regions where molecular composition changes dramatically and where the chemistry is influenced by penetration of cosmic rays or X-rays. 
Following \cite{Jenkins1999a}, for the usual gas-to-dust ratio, the log of the stellar flux at $1150$\,\AA\  decreases by $\sim -6.4\times10^{-22}\times N({\rm H}_{\rm tot}$), thus by $\sim -1.2 \times A_V$ relative to a non-reddened star. An extinction of $A_V=4$ produces an obscuration by a factor $60\,000$. Still, with an effective area three times that of HST/COS, $20$ minutes exposure time could provide a signal-to-noise ratio of $100$ for stars comparable to the bright $\it Copernicus$ targets but obscured by this amount of material.

We list below some of the most important applications enabled by such requirements.

\paragraph{Calibrating the $I_{\rm CO}/N({\rm H}_2)$ relation}

The CO emission is widely used to trace \hmol\ but the CO-to-\hmol\ conversion factor, $X_{\rm CO}$, is known to depend on metallicity and most of the time it relies on indirect measurements of \hmol\ (e.g., \citealt{Bolatto2013a}) such as virial mass, gamma-rays, dust emission, dust absorption, surrogate molecules, all of which with potentially uncertain calibration relationships. By measuring CO and \hmol\ together in absorption in low-optical depths transitions, and CO in emission along the same lines of sight, we can calibrate the CO-to-\hmol\ conversion factor traditionally used for emission lines 
\citep{Burgh2007a,Liszt2008a,Liszt2017a}. By observing lines of sight of different extinction, different metallicity, and resolving the different components in the line of sight,  we can  measure X$_{\rm CO}$ as a function of  cloud depth ($A_V$) and of metallicity (which can then be applied to the low-metallicity systems in the distant universe).

It is also important to characterize the "CO-dark" zones, where hydrogen is already molecular, but carbon still in the form of C$^+$ or C$^0$. These zones, with $A_V= 0.1-1$ \citep[e.g.,][]{Wolfire2010a}, can represent a significant or even dominant amount of H$_2$ (e.g., \citealt{Grenier2005a,Madden1997a}).
\paragraph{How is H$_2$ excited?}

The population of \hmol\ in the  $J>2$ rotational levels in the standard diffuse ISM  is not well understood: is it due to optical pumping radiative excitation \citep{Gillmon2006a} or to the presence of warm \hmol\ \citep{Verstraete1999a,Gry2002a,Falgarone2005a}?
A deeper understanding requires a
better characterization of the gas through the
measurement of temperature and turbulence of \hmol\ in the high-$J$ levels. 
At a resolving power higher than $10^5$, distinctive signatures of individual components or cloud regions with different excitation could be identified. 
Some models invoke shocks or the dissipation of turbulence in vortices to produce the $J>2$ \hmol\ as well
as CH$^+$ \citep{Godard2014a}, with specific signatures in the velocity distribution of the warm \hmol\ gas.

In several cases an  increase in velocity dispersion with $J$ has been observed \citep[often at high redshift because of higher resolution,][]{Noterdaeme2007a,Klimenko2015a}. Towards the star $\zeta$OriA (observed at high resolution with IMAPS) \cite{Jenkins1997a} have interpreted it as \hmol\ being created in a post-shock zone via the formation of H$^-$.
Other interpretations involve energy being transferred to vibrational and rotational excitation upon \hmol\ formation. 
If a significant fraction goes into ejecting the newly formed molecules at large velocity \citep[fast \hmol\ production,][]{Barlow1976a}, \cite{Jenkins1999a} have shown that this would produce detectable broad wings in the high-$J$ \hmol\ profiles, provided the resolution is high enough to differentiate them from the slow \hmol.  Observing the detailed velocity dispersions makes it possible to infer the fraction of the available $4.5$\,eV energy that  is transferred to \hmol\ excitation, to kinetic motion of the molecules, or in the form of heat for the grain. 
\paragraph{Small-scale structures in interstellar clouds}

Knowing both the velocity and spatial structure is critical to describe the \hmol\ self-shielding and the H$^0$-to-\hmol\ transition. 
The spatial structure can be studied through repeated observations of lines of sight as they drift through the foreground clouds due to the motions of the target star or the observer \citep{Boisse2005a,Lauroesch2007a}, for instance through observations of runaway stars or binary stars. One needs to observe atoms or various molecules such as \hmol\ and CO at very high SNR and high resolution, to be combined with such observations of CN, CH, and CH$^+$ in the visible. \cite{Welty2007a} provides an illustration for multi-epoch optical and UV observations of variable absorption in C\1 fine-structure lines tracing variations in local n$_{\rm H}$.
Tracking of spatial and temporal absorption variations  enable a better understanding of the nature and the properties of the so-called tiny-scale atomic structures \citep[$\sim$10$^{1-4}$\,AU,][]{Heiles1997a} thought to be part of a universal turbulent cascade \citep{Stanimirovic2018a}. 
Such observations in the far-UV enable the detection of structures down to potentially much lower H\1\ column densities as compared to H\1\ 21 cm absorption surveys. 

\subsection{Warm gas ionization}

\subsubsection{Ionization structure}

The diffuse H$\alpha$ emission in the Galaxy (which dominates the mass budget of ionized gas) is thought to be related to early-type stars and to supernova-driven turbulence and superbubble structures (e.g., \citealt{Wood2010a}). However, many questions remain regarding (1) some unexpectely large measured temperatures which require heating mechanisms other than photoionization, (2) the spatial distribution of the diffuse gas and the escape fraction of ionizing photons (to be compared to direct derivation of rest-frame Lyman continuum across many lines of sight), and (3) the derivation of useful constraints that can be used for 3D large-scale MHD models (e.g., \citealt{Haffner2009a}).

A better knowledge of the exciting mechanism for the warm diffuse medium requires studying the ionization structure which, in turn, involves getting detailed ionization fractions as a function of ionizing energy. This is possible through the observations of different ionization stages like [S\1, S\2, S\3, S\4, S\6], [O\1, O\6], [C\1, C\2, C\3, C\4], [Si\1, Si\2, Si\3, Si\4], [N\1, N\2, N\3, N\5], and H\1. Many of these species have lines in the UV domain, however important stages only have lines in the far-UV domain. It is therefore imperative to get access to both the UV and far-UV domains at the same high resolution. Locally the simplicity of the short sight lines toward stars in the solar vicinity ($<100$\,pc) provides a unique opportunity to study the ionization structure of individual interstellar regions, clouds, and interfaces, that are usually blended in longer sight lines  \citep{Gry2017a}. The detection of the  weak lines that are critical for these studies requires the possibility to record UV and far-UV spectra of hot nearby stars at high SNR (well in excess of $100$). At larger scales, observing samples of post-asymptotic giant branch (PAGB) and blue horizontal-branch (BHB) stars ($V\sim15$) toward globular clusters that also contain a pulsar (whose dispersion measure yields an integrated value for n$_e$) provides detailed ionization fractions of the WIM as a function of ionizing energy \citep{Howk2012a}.

\subsubsection{Partly-ionized neutral gas}

The ionization fraction and electron density in the neutral gas are important parameters to examine as they tell us about the abundance of free electrons heating the gas and providing pathways for molecular gas formation in the gas phase. They can also be used to infer the influence of supernovae and compact objects on the ISM properties through the propagation of ionizing cosmic rays and soft X-rays. Partly ionized gas can be studied through species with large photoionization cross-section (e.g., N\2, Ar\1) and through the population in fine-structure levels (e.g., the N\2\ multiplet at $1084$\,\AA, C\2*/C\2)\footnote{Absorption lines arising from fine-structure levels are noted with * (** for the second fine-structure level) to differentiate them from the ground-state transitions. }, while temperature can, for instance be determined from Mg\1/Mg\2\ or Si\2*/Si\2\ (e.g., \citealt{Jenkins2000a,Vladilo2003a,Jenkins2013a,Gry2017a}).
High spectral resolution is necessary to disentangle the multiplets while high SNR is necessary to detect weak absorption lines arising from fine-structure levels (e.g., Si\2*, O\1**). 

\subsection{Hot ionized gas}\label{sec:layers}
 
The origin and nature of the collisionally-ionized gas, seen notably in O\6\ absorption in the disk and halo of our Galaxy (corresponding to $T\approx3\times10^5$\,K), is still debated  \citep{Wakker2003a,Savage2006a,Otte2006a,Welsh2008a}: is it formed in radiative cooling  supernovae-shocked gas, in conductive interfaces with cold gas, or in turbulent mixing layers? While information on O\6\ has been limited so far to its bulk column density and its abundance relative to other species, we also need information on its velocity structure and line width in order to infer physical conditions (in particular the temperature) and to disentangle the different components and to relate them to the other low- and high-ionization species (e.g., Si\3, C\4, Si\4, N\5). The electron density can also be calculated through combined observations of O\6\ in absorption and in emission \citep{Otte2006a}. 

The boundaries between the different phases can often be quite abrupt, and it is not yet clear how they trade matter and
entropy. Interfaces between the hot and warm media may play an important role in enhancing the cooling of the hot material through either conductive or radiative losses. Our understanding of the physical properties of such boundaries should help us to construct more accurate accounts on how rapidly hot gas volumes created by supernova explosions dissipate, which, in turn,  influences the morphology of galaxies and some key aspects of the overall cycle of matter and thermal energy within them.

\begin{figure}[h]
\begin{centering}
         \includegraphics[width=0.49\textwidth]{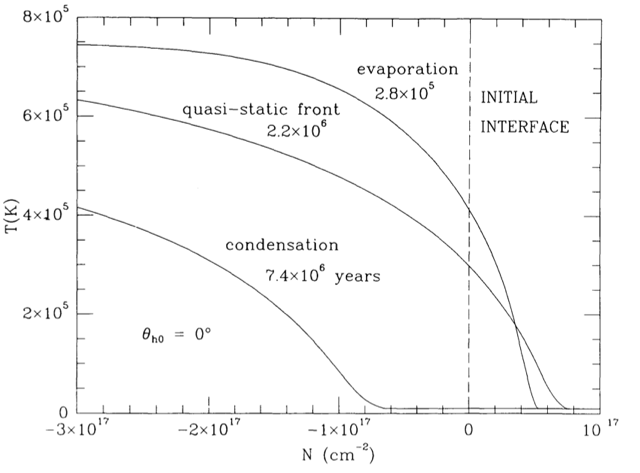}
         \includegraphics[width=0.49\textwidth,height=0.25\textheight]{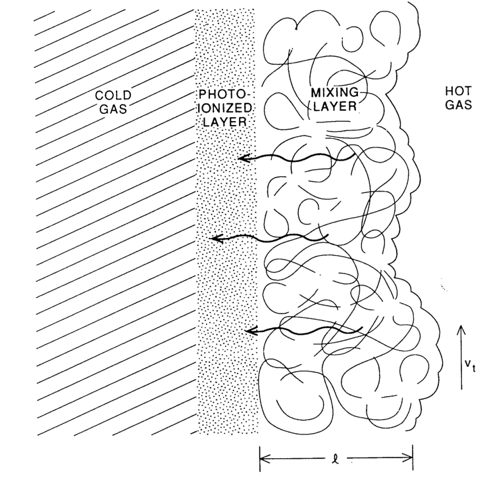}
\caption{\small {\it a)} The structure of temperature vs.\ hydrogen column density in a conduction front. Different curves show changes that occur as the front evolves from a young evaporation front to an older condensation front \citep{Borkowski1990a}. {\it b)} Schematic drawing of a turbulent mixing layer, showing hot and cold gas separated by a thin, intermediate photoionized layer. The hot gas moves at a transverse velocity $v_t$ relative to the layer \citep{Slavin1993a}.}\label{fig:mixing}
  \end{centering}
\end{figure}

Many theorists have confronted this issue and have concluded that two basic categories of interactions can take place: (1) the establishment of a conductive interface where evaporation or condensation can occur \citep{Cowie1977a,McKee1977a,Ballet1986a,Slavin1989a,Borkowski1990a,Dalton1993a,Gnat2010a} (Fig.\,\ref{fig:mixing}{\it a}) and (2) a turbulent mixing layer, where the existence of any shear in velocity between the phases creates instabilities and mechanically induced chaotic interactions \citep{Begelman1990a,Slavin1993a,Kwak2010a} (Fig.\,\ref{fig:mixing}{\it b}), which can ultimately lead to ablation in the extreme cases of the High Velocity Clouds (HVCs) passing through a hot medium \citep{Kwak2011a,Henley2012a}.

Observers have attempted to identify these processes chiefly by analyzing interstellar absorption features of ions that are most abundant at intermediate temperatures, such as Si\4, C\4, N\5 and O\6, and then comparing their column density ratios with theoretical predictions \citep{Spitzer1996a,Sembach1997a,Indebetouw2004a,Zsargo2003a,Indebetouw2004b,Wakker2012a}. In a survey probing the Galactic disk, \cite{Lehner2011a} showed that Si\4\ and C\4\ are found in both broad and narrow components, and the high-ion column density ratios exhibit substantial variations in most lines of sight,
which implies that very different processes operate in different environments in the Galaxy. However, the confusion caused by the overlap of many different regions  over the large distances covered by surveys (e.g., for O\6, \citealt{Jenkins1978a,Jenkins1978b,Bowen2008a}) has made it difficult to get
a clear picture of the nature of these interfaces.

The possibility of much simpler lines of sight is offered within the local ISM, where a single warm, diffuse cloud accounts for most of the matter within the first $50$\,pc \citep{Gry2014a}, generating a single interface with the surrounding hot bubble gas.
Its signature should be observed in C\4, N\5, O\6, Si\3, Si\4\  toward hot nearby stars (white dwarfs or B stars). Up to now the detection of this interface has been elusive, mostly due to the extreme weakness of the expected absorption -- SNR in excess of $200$ is necessary. Such SNR is in principle easy to reach with nearby hot stars, but it has often been limited by the bright-object limits of UV detectors and drastic neutral density filters.

Spectral resolution is also a key requirement to understand the highly ionized gas and its relation with other gas-phases. This is well illustrated by the study of \cite{Lehner2014a} who, thanks to an improved spectral resolution by a factor of $\approx2$ enabled by instruments in the visible as compared to rest-frame studies with FUSE, uncovered differences between the profiles of [O\6, N\5] and [Si\4, C\4] suggesting that the bulk of the O\6\ absorption is produced in a radiatively cooling gas produced between a shock-heated outflowing gas rather than in a very diffuse, extended gas photoionized by the extragalactic UV background radiation (see also Sect.\,\ref{sec:cgm}).

Last but not least, high SNR spectra obtained toward stars can reveal weak components at high velocities, caused by cooling layers behind shock fronts \citep{Welty2002a}. Tracking velocity shifts and line widths for ions that should appear at different locations in downstream flows offers insights on how the gas cools and recombines in a time-dependent ionization scheme. The depth of absorption features in radiative shocks is expected to be larger than a few \%\ in S\3, Si\4 and C\4\ for large enough shock velocities, and should be detectable in spectra recorded at SNR in excess of a few hundreds.

\subsection{Dynamics in the diffuse multiphasic turbulent interstellar gas}\label{sec:dynamic}

In many ways, the latest generation of astronomical instruments has shaken our understanding of the diffuse interstellar matter. Long thought to be composed of independent phases which could be studied in the framework of static models at chemical equilibrium, the diffuse interstellar gas is in fact restless, turbulent and magnetized, as well as out-of-dynamical, -thermal, or -chemical equilibrium.

Because of thermal instability, the neutral gas is known to settle in two stable phases, the WNM and the CNM which are roughly at thermal pressure equilibrium. However, theoretical predictions (e.g., \citealt{Hennebelle2007a}) as well several sets of measurements complicate this simple picture. (1) The discovery for the past 15 years of the Lukewarm Neutral Medium (LNM) phase observed in H\1\ shows that as much as $30$\%\ of the gas mass is in the unstable regime (e.g., \citealt{Kalberla2018a,Marchal2019a}), indicating that the exchange of matter between the WNM and the CNM is paramount. (2) The gas pressure deduced from C\1\ UV absorption lines (e.g., \citealt{Jenkins2011a}) shows a strong dispersion in the local ISM; whether this dispersion is due to variations in the local irradiation conditions, the action of turbulence, the presence of self-gravitating clouds along the lines of sight, or a combination of all, is still an ongoing issue.
(3) The infrared dust emission and its polarization indicates that the CNM is organized in projected filamentary structures whose orientations and physical conditions are tightly linked to the orientation and strength of the local magnetic field (e.g., \citealt{PlanckCollaboration2018a}). (4) The characterization of the kinematic and chemical signatures of molecules and
molecular ions with \textit{Herschel} and SOFIA finally suggest that the different phases are entangled in the production and excitation of chemical species in the diffuse gas (e.g., \citealt{Neufeld2015a}).

On the theoretical side, astrochemical models have drastically improved during the past decade to study the at- or out-of-equilibrium chemistry in 3D dynamical models of isothermal turbulence (e.g., \citealt{Bialy2017a}), monophasic CNM turbulence at
intermediary (e.g., \citealt{Glover2010a}) or dissipative scales (Lesaffre et al.\ submitted), and multiphasic turbulence at intermediary (e.g., \citealt{Levrier2012a,Valdivia2016a,Valdivia2017a}) or galactic scales \citep{Seifried2017a,Girichidis2018a}. It is concluded that the combination of turbulence, magnetic field, and gravity strongly perturbs the chemical composition of the diffuse matter though density fluctuations induced by supersonic motions, out-of-equilibrium mixing at phase interfaces, or dissipation processes. The H$^0$-to-H$_2$ transition, the CO/H$_2$ abundance ratio, and the abundances and kinematics of atomic and molecular tracers are all affected.

Several questions arise from these studies that can only be answered through the systematic observations of a statistically significant number of sources at several wavelengths, and most notably in the UV range. What are the mass transfer rates between
the different phases of the ISM? What are their survival timescales and their volume filling factor? How far from equilibrium are the ionization and molecular fractions? What fraction of H$_2$ belongs to the LNM thermally unstable phase? What are the separate
roles of supersonic motions, turbulent dissipation, and turbulent transport in the chemical richness observed in the ISM ? Finally, what does a statistical study of the ionization and molecular fractions tell us about the magnetic field and the thin Faraday structures observed with LOFAR (e.g., \citealt{Zaroubi2015a,VanEck2019a})?

In this context, a far-UV spectrograph is necessary to get access to the amount of H$^0$, H$_2$, and of atomic ions in the local interstellar gas (in a radius of $4$\,kpc around the Sun). High spectral resolution observations are mandatory to extract the kinematics, and hence to identify the phases responsible for a given absorption profile. High sensitivity will allow a survey over hundreds of targets, including short and long lines of sight with high extinction, which can be used in synergy with GAIA, \textit{Planck}, and SKA data to build a statistical sample of the local interstellar diffuse medium that can be compared with state-of-the-art dynamical models.

 \subsection{Detection of very faint lines of scarce elements}
 
A number of scarce elements with high astrophysical interest require very high SNR to be detected because of the weakness of their lines, most of which being in the far-UV domain. Let's mention: (1) the light elements like $^{10}$B, $^{11}$B \citep{Lambert1998a}, $^6$Li, $^7$Li (in the visible; \citealt{Meyer1993a}), or Be \citep[undetected,][]{Hebrard1997a}; (2) The r- and s- process elements like Ga, Ge, As, Se, Kr, Cd, Sn, Pb \citep{Ritchey2018a}. It would be interesting to look for localized enhancements of these elements in regions (serendipitous discovery!) where a neutron star merger occurred in a time that is more recent than a mixing time for the ISM; and (3) Isotope ratios of atomic and molecular species. For instance HCl, whose line has barely been detected at $1290$\,\AA, may be split in lines of H$^{35}$Cl and H$^{37}$Cl. However the interpretation of molecular isotopes may be confused by the influence of chemical reactions. Atomic shifts are of order of a few \kms\ in some cases, so measurements should be feasible with high resolution in cold regions where velocity dispersions are low.

\section{Precision measurement of magnetic field from near to far, from small to large scales in ISM}\label{sec:bfield}

\textbf{
\begin{itemize}
\item \textit{What is the role of the magnetic field in the distribution of ISM phases? How does the magnetic field modulate the transport of matter for star formation?}
\item \textit{What impact does ground state alignment have on the physical parameter derivation (abundances, ionization...)?}
\end{itemize}}

\bigskip

Magnetic fields have important or dominant effects in many areas of astrophysics, but have been very difficult to quantify. Spectropolarimetry from Ground State Alignment (GSA) has been suggested as a direct tracer of magnetic field in interstellar diffuse medium. The alignment itself is an effect well studied in the laboratory: the effect arises due to the ability of atoms/ions with fine and hyperfine structure to get aligned in the ground/metastable states. Owing to the long life of the atoms on ground states, the Larmor precession in an external magnetic field imprints the direction of the field onto the polarization of absorbing species. This provides a unique tool for studies of sub-gauss magnetic fields using polarimetry of UV, optical and radio
lines. Many spectral lines with strong signals from GSA are in the UV band. By discerning magnetic fields in gas with different dynamical properties, high spectral resolution measurement of spectral polarization will allow the study of 3D magnetic field distribution and interstellar turbulence.

GSA also provides a unique chance to map 3D direction of magnetic field on small scales, e.g., disks, where grain alignment is unreliable. The range of objects suitable for studies is extremely wide and includes magnetic fields in the interplanetary medium, in the ISM, and in circumstellar regions as well as diffuse media in extragalactic objects. Last but not least, the consequences of the alignment should be taken into account for correct determination of the abundances of alignable species.

\subsection{Magnetic field measurement in the ISM}

Magnetic fields play a crucial role in various astrophysical processes, including star and planet formation, accretion of matter, or transport processes. Recent dust polarization measurements by, e.g., \textit{Planck}, have represented a huge step forward in the knowledge of the Galactic magnetic field in terms of sensitivity, sky coverage, and statistics (e.g., \citealt{PlanckCollaboration2018a}). However, they unfortunately tell us nothing on the distance of the magnetic field and its distribution in the different components or phases. Furthermore, they can only provide the integrated mean orientation of the magnetic field projected onto the plane of the sky, as they cannot quantify the strength of the magnetic field nor access the third dimension. 

Therefore it is important to explore new effects which can bring information about the magnetic field properties and signatures. ``Ground state alignment'' has been identified as an innovative way to determine the magnetic field in the diffuse medium. The atoms get aligned in terms of their angular momentum and, as the life time of the atoms/ions we deal with is long, the alignment induced by anisotropic radiation is sensitive to very weak magnetic fields ($10^{-15}-1$\,G, \citealt{Yan2012a}), which is precisely the level of the magnetism in diffuse medium, including in both ISM and IGM. It must be noted that even the general interstellar radiation field presents enough anisotropy to align the atoms and create GSA \citep{Zhang2015a}. Observing several hundred hot stars with a SNR of $500$ to measure linear polarization from optically thin UV absorption lines will provide us exclusive information on magnetic field distribution and turbulence properties in different interstellar phases.

\textit{Most of the resonance absorption lines are in the UV domain}. A UV band polarimeter with high spectral resolution ($R>20\,000$) will thus provide an incomparable opportunity for precision magnetic field measurement, which no other current instruments can offer. Particularly, the high spectral resolution allows simultaneous determination of both velocity and magnetic field, filling the gap of 3D magnetic tomography in ISM, which is so far missing. The resonance absorption lines appropriate for studying magnetic fields in diffuse, low column density ($A_V \sim$ few tenths)
neutral clouds in the interstellar medium are those from N\1, O\1, S\2, Mn\2, and Fe\2, all in the UV range. At higher column densities, the above lines become optically thick, and excited states become available as well as lines from less abundant species.

 \begin{figure*}[h]
\begin{centering}
         \includegraphics[width=0.3\textwidth]{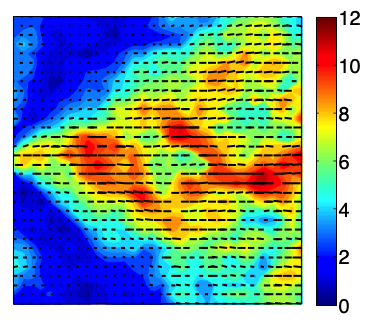}
         \includegraphics[width=0.3\textwidth,height=0.25\textheight]{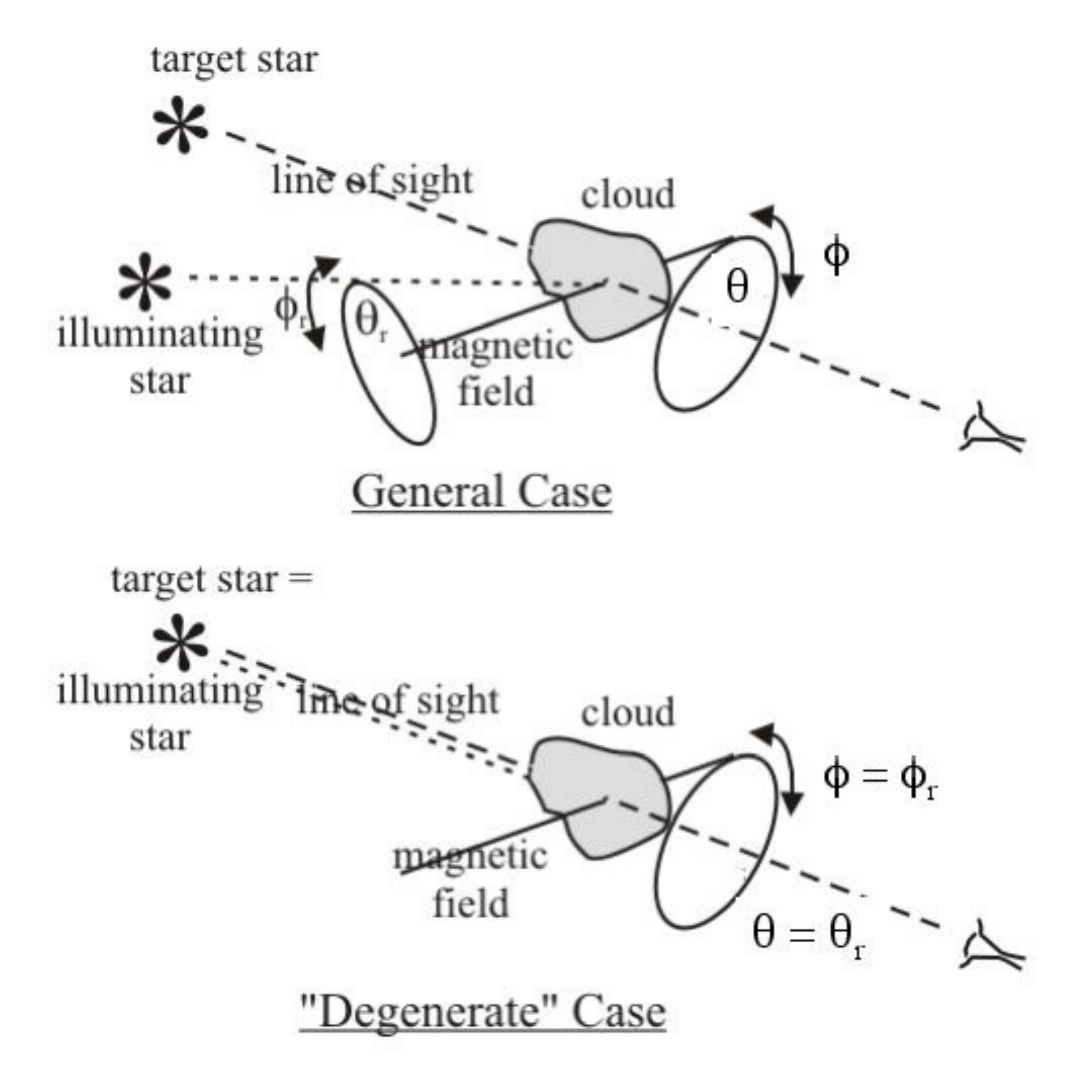}
         \includegraphics[width=0.3\textwidth,height=0.25\textheight]{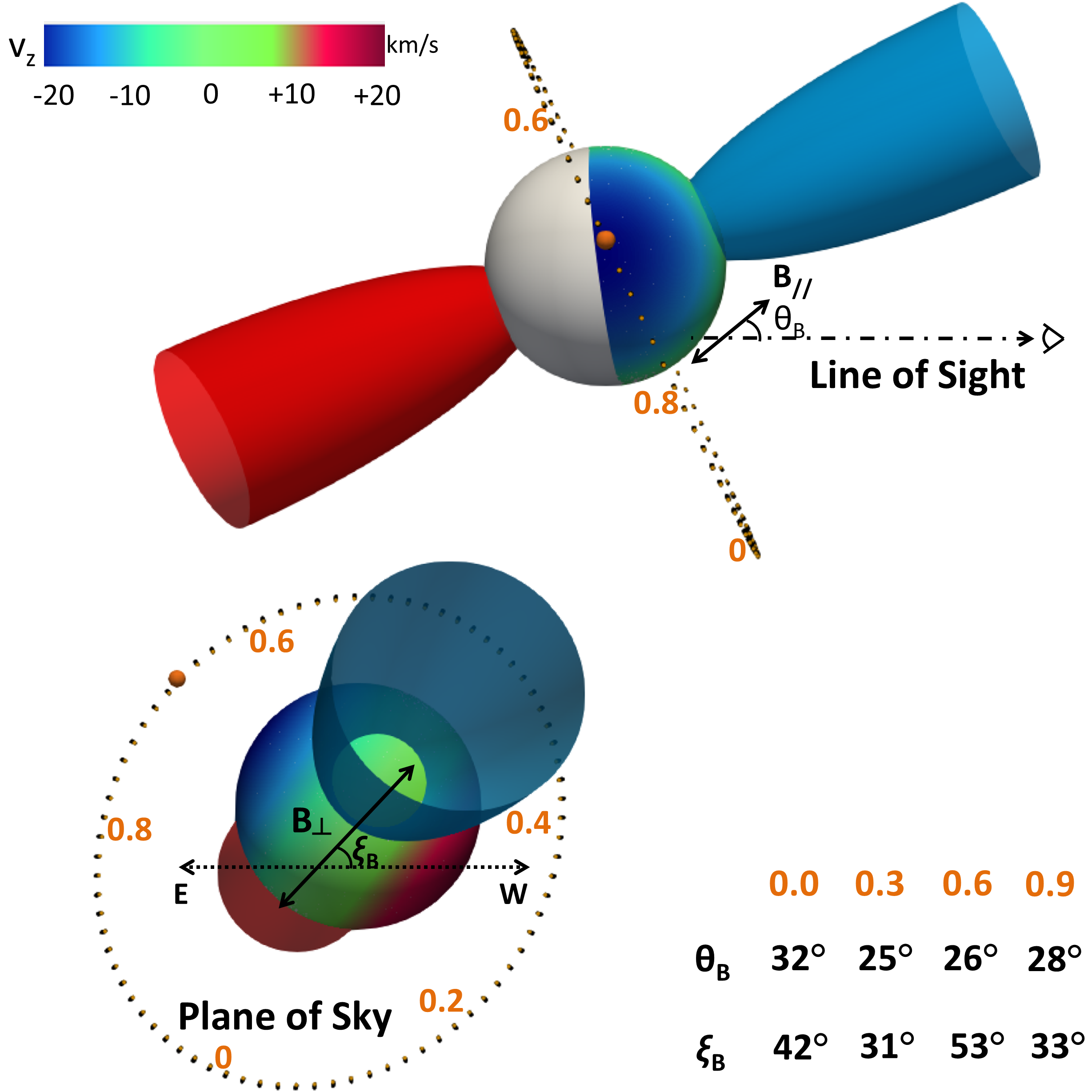}
\caption{\small {\it a)} Synthetic polarization map of the simulated super-Alfvenic diffuse ISM. The size of the field is $1$\,pc$^2$, with an O/B star located $0.1$\,pc to the left. (a) The contour color reveals the percentage of polarization induced in the S\2\ $1250$\,\AA\ absorption line and the orientation of the bars represents the direction of the polarization. The expected polarization is mostly above $5$\%. 
{\it b)} Typical astrophysical environment where 
GSA can occur. A pumping source deposits angular momentum to 
atoms in the direction of radiation and causes differential occupations on 
their ground states. In a magnetized medium where the Larmor precession 
rate $\nu_L$ is larger than the photon arrival rate $R_F$, however, 
atoms are realigned with respect to magnetic field. Observed polarization depends on both $\theta_r$ (angle between magnetic field and illuminating star) and $\theta$ (angle between the magnetic field and the line of sight). 
In general, there are two situations: the alignment 
is produced by a pumping source while we observe another weak background 
source whose light passes through the aligned medium ({\it upper part}) or the background source coincides with the pumping source, in which case $\theta_r=\theta$ ({\it lower part}). {\it c)} 3D topology of magnetic field in the 89\,Her post-AGB binary system. The system is plotted from two different orientations showing the line-of-sight and plane-of-sky projections. The color scale indicates the line-of-sight velocity ($v_z$) of the medium. The plane-of-sky projection of the symmetric axis of the outflow is $45^\circ$ to the East-West direction. The inferred 3D magnetic field directions for different orbital phases are displayed \citep{Zhang2019a}.
}\label{fig:gsa}
  \end{centering}
\end{figure*}

As a first step, with low resolution measurement, 2D magnetic field can be easily obtained from the direction of polarization with a $90^\circ$ degeneracy, similar to the Goldreich-Kylafis effect \citep{Goldreich1981a} for molecules and similar to grain alignment according to some theories. 

UV absorption lines are polarized through GSA exclusively. Any polarization, if detected, in absorption lines, would not only be an exclusive indicator of alignment, but also of the magnetic field since no other mechanisms can induce polarization in absorption lines. With a high resolution spectropolarimeter, 3D direction of magnetic field can be extrapolated by combining the polarization of two lines or one line polarization with their line intensity ratio, which is influenced by the magnetic field as well \citep[see][]{Yan2006a,Zhang2018a}. With the knowledge of the degree of polarization and/or $\theta_r$, the angle between magnetic field and line of sight, the $90^\circ$ degeneracy can be lifted.

\subsubsection{Large scale magnetic field distribution and turbulence: 3D tomography}
 
On large scales, spectropolarimetric maps must constrain much better the 3D distribution of the magnetic field. The interstellar magnetic field is turbulent with velocity and magnetic fluctuations ranging from large injection scales to small dissipation scales. High resolution spectroscopy and spectropolarimetry combined bring forth a wealth of information on interstellar turbulence. Most magnetic diagnostics only constrain averaged mean magnetic field on large scales. In this respect GSA fits a unique niche as it reveals the small scale structure of magnetic field (see Fig.\,\ref{fig:gsa}{\it a}). Measuring 3D turbulence will shed light on many open questions regarding for instance star formation and cosmic ray or interstellar chemistry (see, e.g., \citealt{Hennebelle2019a}).

In highly turbulent environments, we expect magnetic fields to be entangled. A UV instrument with sufficient spectral resolution would be valuable since it naturally reduces line of sight averaging. If the pumping star is along the line of sight, as in the central star of a reflection nebula, this is the so-called ``degenerate case'' (see Fig.\,\ref{fig:gsa}{\it b}), where the position angle of the polarization provides the 2D magnetic field in the plane of sky, and the degree gives the angle to the line of sight. In the more general case, though, an observed cloud might be pumped from the side and the positional angle of magnetic field is available with $90^\circ$ degeneracy, so that the derivation of the full magnetic geometry requires measuring two lines (not necessarilly from the same species). 
  
\subsubsection{Small scale magnetic field: 3D direction}

On small scales, spectropolarimetry from GSA is an ideal tracer of local magnetic fields. Examples include disks, local bubble, and PDR regions. One interesting case is that of circumstellar disks, for which grain alignment has been found unreliable. In the case of pre-main sequence stars, pumping conditions are similar to those for comets in the Solar System  \citep[see][]{Shangguan2013a}: pumping rates on the order of $0.1-1$\,Hz, and realignment for fields greater than $10-100$\,mG. Conditions here seem to be conducive to substantial populations in CNO metastable levels above the ground term: \cite{Roberge2002a} find strong absorption in the FUV lines ($1000-1500$\,\AA) of O\1\ (1D) and N\1\ and S\2\ (2D), apparently due to dissociation of common ice molecules in these disks (also common in comet comae). Since these all have total angular momentum quantum number $>1$, they should be pumped and realigned. This presents the exciting prospect of detecting the magnetic geometry in circumstellar disks and monitoring them with time. The potential has been clearly revealed by the detection on a binary system where 3D magnetic fields are precisely mapped for the first time via polarization of absorption lines (see Fig.\,\ref{fig:gsa}c; \citealt{Zhang2019a}).

\subsubsection{Polarization of other lines}

\paragraph{Resonance and fluorescence lines} The magnetic realignment diagnostic can also be used in resonant and fluorescent scattering lines.  This is because the
alignment of the ground state is partially transferred to the upper
state in the absorption process \citep{Yan2007a}.  If the direction of optical pumping is known, e.g., in planetary system and circumstellar regions, magnetic realignment induces a line polarization whose positional angle is neither perpendicular or parallel to the incident radiation. This deviation depends on the magnetic geometry and the scattering angle.  The degree of polarization also depends on these two factors. In practice, GSA can be identified by comparing the polarizations from non-alignable (which do not trace the magnetic field) and alignable species. There are many fluorescent lines in emission nebulae that are potential candidates \citep[see][]{Nordsieck2008a}. Reflection nebulae would be an ideal place to test the diagnostic, since the lack of ionizing flux limits the number of levels being pumped, and especially since common fluorescent atoms like N\1\ and O\1\ would not be ionized, eliminating confusing recombination radiation.

\paragraph{IR/submillimeter transitions within ground state} The alignment on the ground state affects not only the optical or UV transitions to the excited state, but also the magnetic dipole transitions within the ground state. The submillimeter lines emitted and absorbed from the aligned medium are polarized in a same fashion as that of the absorption lines, i.e., either parallel or perpendicular to the magnetic field \citep{Yan2008a,Zhang2018c}. Current facilities, e.g., SOFIA and ALMA, already have the capability to cover the submillimeter band for the spectral polarimetry observation. One can also mention the linear polarization of the $21$\,cm line (e.g., \citealt{Clark2019b}). 

\subsubsection{Magnetic field strength}

GSA is usually by itself not directly sensitive to the magnetic field strength. The exception is a special
case of pumping photon absorption rate being comparable with the Larmor frequency \citep[see][]{Yan2008a,Zhang2019a}. However, this should not
preclude the use of GSA for studies of magnetic field. Grain alignment is not sensitive to magnetic field strength either.
This does not prevent polarization arising from aligned grains to be used to study magnetic field strength with the so-called
Chandrasekhar-Fermi technique \citep{Chandrasekhar1953a}. In this technique the fluctuations of the magnetic field direction are associated with Alfven perturbations and therefore simultaneously measuring the velocity dispersion using optical/absorption lines arising from the same regions
it is possible to estimate the magnetic field strength. The Chandrasekhar-Fermi technique and its modifications (see, e.g., \citealt{Hildebrand2009a,FalcetaGoncalves2008a,Cho2016a}) can be used to derive the magnetic field strength using GSA. 

The advantage of using spectral lines compared to dust grains is that both polarization and line broadening can be measured from the same lines, making sure that both polarization and line broadening arise from the same volumes. In addition, GSA, unlike grain alignment does not contain ambiguities related to dust grain shape. Thus, potentially, Chandrasekhar-Fermi technique can be more accurate when used with GSA. 
 
\subsection{Synergy of techniques for magnetic field studies}

The GSA and grain alignment complement each other. For instance, measurements of grain alignment in the region where GSA is mapped for a single species
can remove ambiguities in the magnetic field direction. At the same time, GSA is capable of producing a much more detailed map of magnetic field in the diffuse gas and measure magnetic field direction in the regions where the density
of dust is insufficient to make any reliable measurement of dust polarization. In addition, for interplanetary magnetic field measurements it is important that GSA can measure magnetic fields on time scales much shorter than aligned grains. 

The synergy exists with other techniques as well. For instance, GSA allows to reveal the 3D direction of magnetic field. This gives the direction of the magnetic field, but not its magnitude. If Zeeman effects allow to get the magnitude of one projected component of magnetic field, this limited input enables GSA to determine the entire 3D vector of magnetic field including its magnitude. Such synergetic measurements are crucial.

As astrophysical magnetic fields cover a large range of scales, it is important to have techniques to study magnetic
fields at different scales. For instance, we have discussed the possibility of studying magnetic fields in the interplanetary medium. This can be done without conventional expensive probes by studying polarization of spectral lines. In some cases
spreading of small amounts of sodium or other alignable species can produce detailed magnetic field maps of a particular regions of the interplanetary space, e.g., the Earth magnetosphere.

\section{Beyond the Milky Way: the ISM properties in external galaxies} \label{sec:ism}

\textbf{
\begin{itemize}
\item \textit{How can we remove biases in abundance determinations in unresolved nearby galaxies? Can we map the ecosystem of interstellar clouds in external galaxies, is there evidence of metal-free gas accretion?}
  \item \textit{Does star formation proceed in cold atomic gas in quasi-pristine environments at redshift $\sim0$? What is the influence of compact objects in the ISM properties and on star formation?}
\end{itemize}}

\bigskip

While past and present UV spectroscopic instruments have greatly improved our knowledge of the ISM physics and chemistry in the Milky Way, the extragalactic ISM is mostly uncharted territory with these instruments. Much of what we know is based on near-UV to far-infrared spectroscopic observations. Only a limited number of nearby galaxies could be investigated in the far-UV, with unavoidable biases concerning the confusion of lines of sight and concerning the low required spectral resolution owing to the low fluxes. It is now urgent to reach the same level of details on the ISM properties for nearby galaxies as what is currently possible for the Milky Way.

\subsection{Solving column density determination biases}\label{sec:biases}

In the far-UV, apart from a few lines of sight toward individual stars in the Magellanic Clouds (e.g., \citealt{Welty2016a,RomanDuval2019a}), the extragalactic ISM has been mostly observed toward stellar clusters and at low spectral resolution ($R\lesssim20\,000$), with inherent biases for the column density determination. First, unresolved absorption lines observed toward a single line of sight may show a low apparent optical depth (i.e., corresponding to the linear regime of the curve of growth) even though some individual velocity components are saturated, leading to the ``hidden'' saturation problem and to the possible underestimate of column densities by factors of a few or more (e.g., \citealt{James2014a}). Second, multiple lines of sight toward stars of different brightness may contribute to the observed spectrum, with each line of sight intersecting multiple interstellar clouds with different properties (metallicity, column density, turbulent velocity, radial velocity). The resulting combination is highly non-linear, especially if some of the individual (i.e., a given line of sight and a given cloud) components are saturated.

While the biases related to saturated components can be mitigated for a single line of sight with a suite of transitions of varying strengths and with a well-behaved distribution of components \citep{Jenkins1986a}, biases related to the multiplicity of lines of sight have been little explored, even in the favorable case of unsaturated components, notably because the spatial distribution of bright stars in the dispersion direction of slit spectrographs complicates even further the line profile and its analysis (e.g., \citealt{Lebouteiller2006a}).  

Hence robust column density determinations in nearby galaxies ideally require a spectral resolution high enough to disentangle $\approx2$\,\kms\ wide components (typically observed in the Milky Way) and a spatial resolution high enough to resolve individual stars. The corresponding signal-to-noise requirement quickly becomes prohibitive for galaxies further than a few Mpc but satisfactory compromises can be obtained by observing (1) nearby stellar clusters -- such as those observed with HST or FUSE -- with improved spectral resolution $R\sim10^5$, (2) distant/faint stellar clusters with $R\sim20\,000$, i.e., similar to most current extragalactic ISM spectra, or (3) individual O/B stars with $R\sim20\,000$. Another alternative, as proposed by \cite{Kunth1986a}, is to use background QSOs (to be identified with \textit{Euclid}, LLST...), with similar spectral resolution requirement, provided they are bright enough in rest-frame far-UV where absorption lines in nearby galaxies are probed (e.g., \citealt{Bowen2005a,Kozlowski2013a}).

Overall, an observatory versatile enough to propose high spatial and spectral resolution would solve most of the systematic issues in deriving robust column densities in external galaxies using stellar clusters as background light, thereby nicely complementing studies of Damped Lyman-$\alpha$ systems (DLAs; \citealt{Wolfe2005a}). We review in the following the corresponding science motivations.

\subsection{Pristine gas accretion?}\label{sec:chemab}

Complex lines of sight and limited sensitivity have mostly restricted the study of the extragalactic ISM in the far-UV to resonance lines and chemical abundance determinations, although fine-structure lines can be observed in some nearby galaxies (Sect.\,\ref{sec:sf}). \cite{Kunth1994a} proposed a method to measure neutral gas abundances in blue compact dwarf galaxies (BCDs) using unresolved massive stellar clusters in H\2\ regions as background continuum, thereby providing independent results from abundances traditionally derived using optical emission lines arising in the ionized gas of H\2\ regions. The comparison led to a still ongoing debate, which is well illustrated by several studies of the BCD I\,Zw\,18 ($18$\,Mpc, $\approx2\%$ solar metallicity). Early observations with HST/GHRS showed a discrepancy between the oxygen abundance measured in emission and in absorption, leading to the hypothesis of self-enrichment by the current starburst episode \citep{Kunth1986a,Kunth1994a} but, due to a limited sensitivity, only strong lines were accessible and hidden saturation could not be identified easily \citep{Pettini1995a}. Later studies of the same galaxy with FUSE, which enabled the observation of several metal lines and of hydrogen, highlighted issues regarding the stellar continuum and the selection of weak lines \citep{Aloisi2003a} and showed that only a small discrepancy may exist, if any \citep{Lecavelierdesetangs2004a}. More recently, \cite{Lebouteiller2013a} confirmed, using HST/COS and weak lines such as $\lambda1254$ S\2, that a small discrepancy does exist in I\,Zw\,18. Overall, studies have showed that weak lines (also $\lambda1356$ O\1) may minimize column density determination biases, but at the expense of an in-depth analysis of abundance ratios and of the spatial distribution of metals.

\begin{figure}
  \includegraphics[width=0.5\textwidth]{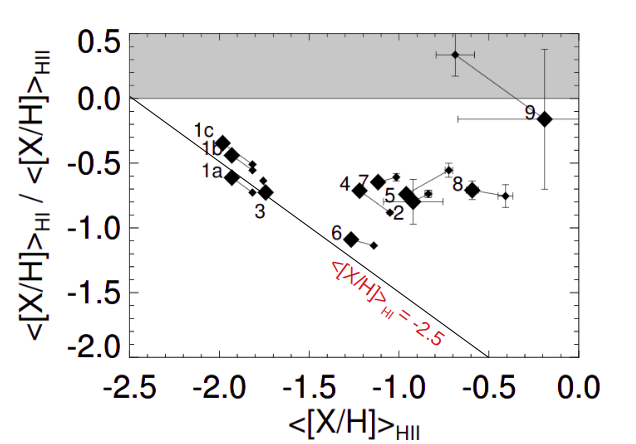}
  \caption{\small Abundance discontinuity between the neutral (H\1\ region; observed in absorption with FUSE and HST) and ionized (H\2\ region; observed in emission) phases in a sample of BCDs \citep{Lebouteiller2009a,Lebouteiller2013a}. Two different methods (diamonds) are compared for each object (numbers). Globally neutral abundances are lower by a factor of a few and the minimum metallicity derived in the neutral phase is around $-2.5$\,dex. }\label{fig:chemab}
\end{figure}

Such discrepancies are important to quantify robustly to understand the distribution of metals and metallicity buildup in galaxies. A sample analysis of neutral gas abundances in BCDs by \cite{Lebouteiller2009a} showed that an overall metallicity floor of $\sim2\%$ solar may exist for galaxies in the nearby Universe which could be linked to the IGM enrichment (Fig.\,\ref{fig:chemab}). 
Metallicity discontinuity between the ionized and neutral phases seem to occur preferentially for moderately metal-poor ($10-50\%$ solar) galaxies, which could be due to dilution by metal-poor/free gas in the halos rather than by self-enrichment in the H\2\ regions (see also \citealt{James2015b}). It is worth noting that extremely low metallicities ($\lesssim1$\% solar) in CGM absorbers at $0.1 \lesssim z \lesssim 1$ are preferentially found in low H$^0$ column densiy clouds ($N({\rm H})<10^{18.5}$\,cm$^{-2}$; e.g., \citealt{Lehner2013a,Lehner2019a,Wotta2016a,Wotta2019a}), which raises the question of the role and properties of infalling clouds in the observed metallicity discontinuity in some BCDs. It is now urgent to perform a complete census of the metals in and around low-mass galaxies in order to understand star formation in metal-poor environments. 

The presence of quasi pristine gas in the outskirts of galaxies has important implications for the galaxy evolution (e.g., infalling scenarii, dispersal/mixing of elements). For instance, the dust-to-gas mass ratio (DGR) obtained for nearby galaxies shows a steep dependence with metallicity \citep{RemyRuyer2014a} which is at variance with the shallower trend obtained for DLAs (see \citealt{Galliano2018b}). On the one hand, metallicity in nearby galaxies was derived in the ionized gas near young star-forming regions where most of the emitting dust is presumably located. On the other hand, the relative dust abundance in DLAs is derived from the depletion strength of refractory species in the far-UV and the metallicity is derived from lines of sight intersecting the entire galaxy body, also in the far-UV, \textit{including regions that may be quasi-pristine or pristine (i.e., metal and dust-free)}. Hence the observed DGR vs.\ metallicity slope in DLAs could reflect a dilution factor, with the dust-rich regions having properties in fact similar to the Milky Way value. High spectral resolution is necessary to decompose the velocity profile in metal lines, to infer the corresponding and expected H\1\ absorption, and to compare to the
observed one in order to quantify the dilution factor.

It should be also noted that the determination of elemental abundances in nearby metal-poor galaxies also provide a powerful tool to understand depletion patterns (dust composition) and strengths (dust-to-metal ratio) as a function of metallicity, but such a technique is limited by the small number of metal-poor galaxies with far-UV absorption spectra (see \citealt{RomanDuval2019b}). The abundance of deuterium also needs to be explored in the metal-poor ISM, either through D\1/H\1, HD/\hmol, or D\1/O\1\ (the two latter minimizing systematic effects when comparing lines on very different part of the curve of growth; \citealt{Hebrard2002a}), in order to mitigate astration effects (destruction in interiors of stars) and to provide potentially better constraints for Big Bang nucleosynthesis models.

A significant gain in sensitivity is now required to obtain a large sample of metal-poor galaxies (for instance drawn from SDSS; \citealt{Izotov2019a}) while a gain in sensitivity and spatial resolution is required to target individual stars in nearby low-metallicity systems (e.g., Leo\,P, $1.6$\,Mpc, $3\%$ solar; \citealt{McQuinn2015b}) with expected continuum fluxes around $10^{-16}$\,erg\,s$^{-1}$\,cm$^{-2}$\,\AA$^{-1}$. In addition, spectral and spatial resolution are required to solve various biases regarding column density determinations (Sect.\,\ref{sec:biases}) and to determine exact spatial origin of the absorption within the galaxy, which is still unknown. On the other hand, abundances derived from optical emission-lines also suffer from some systematic uncertainties, with a discrepancy observed between abundances derived from collisionally-excited lines and recombination lines (e.g., \citealt{Esteban2002a,Esteban2016a}). This particular discrepancy needs to be explored further by accessing faint recombination lines and abundances in the photosphere of young stars as comparison for various metallicities (e.g., \citealt{Bresolin2016a}).

\subsection{Gas reservoirs for star formation}\label{sec:sf}

While most far-UV absorption lines observed in nearby galaxies toward stars or clusters correspond to resonance lines of atomic species, fine-structure atomic transitions and molecular transitions have been detected in a few objects, paving the way to a better understanding of the star-forming gas reservoir. The apparent lack of molecular gas in nearby star-forming low-metallicity galaxies (e.g., \citealt{Taylor1998a,Cormier2014a}) poses fundamental questions regarding the exact role of H$_2$ in the star-formation process as compared to the more generally defined cold dense gas (e.g., \citealt{Glover2012a}). While CO is often used to trace H$_2$, it is expected that CO emission is globally weaker in low-metallicity galaxies because of lower C and O abundance and because of selective photodissociation of CO in a dust-poor environment (e.g., \citealt{Wolfire2010a,Schruba2012a}), leading to a potentially large or even dominant reservoir of ``CO-dark'' H$_2$ gas \citep{Grenier2005a}. Accessing both CO and H$_2$ absorption lines would allow a direct measurement of the CO-to-H$_2$ conversion as a function of metallicity and extinction (from translucent to truly molecular), a conversion that is notoriously uncertain. At the same time, other molecules such as OH, CH$^+$, or HD could also be examined as potential tracers of the CO-dark H$_2$ gas.  

Accessing H$_2$ in absorption in molecular clouds is the most direct way to probe molecular gas in low-metallicity environments but it has been limited to translucent clouds ($A_V\sim1-3$), with the lack of diffuse H$_2$ detections in the far-UV in the metal-poor ISM (e.g., \citealt{VidalMadjar2000a}) being explained by enhanced photodissociation and a larger critical surface density for H$_2$ formation \citep{Hoopes2004a,Sternberg2014a}. As observations of molecular clouds in low-metallicity galaxies reach smaller spatial scales, in particular with ALMA, it seems that H$_2$ may exist mostly in dense clumps of size $\lesssim 1$\,pc in such environments (e.g., \citealt{Rubio2015a}, Shi et al.\ in preparation). Such clumps may be identified thanks to near-infrared observations of warm H$_2$ layers (e.g., \citealt{Thuan2004a,Lebouteiller2017a}) but the determination of physical properties (temperature, density, magnetic field, DGR) as a function of the environment (e.g., Milky Way vs.\ low-metallicity galaxies, quiescent vs.\ active star-formation) requires the observation of H$_2$ absorption lines in various rotational and vibrational levels. 

Finally, thermal processes can be investigated through the use of absorption lines arising from the fine-structure levels such as C\2*, O\1*, O\1**, Si\2*... Such tracers give valuable information on the ionization degree, temperature, and density of the neutral star-forming gas reservoir and provide indirect constraints on the gas heating mechanisms (photoelectric effect on dust grains, ionization by far-UV or X-ray photons, shocks...) that are independent and complementary to the information provided by far-IR cooling lines. Fine-structure absorption lines have been observed in and around the Milky Way, in DLAs (shifted to the optical domain), and a few nearby BCDs (e.g., \citealt{Lehner2004a,Wolfe2003a,Howk2005a,Lebouteiller2013a,Lebouteiller2017a}) but the number of Si\2*\ and O\1**\ detections (required to measure the gas temperature) remains small due to limited sensitivity. Through fine-structure cooling lines in absorption one can hope to measure the thermal balance in the gas in regions of various extinctions well resolved in space as compared to IR observations, including in low column density infalling filaments/clouds.

\subsection{Nature of compact objects in primitive environments and their influence on the ISM}\label{sec:compact}

The presence of energetic X-ray binaries may influence the ISM properties in galaxies with extremely low DGR and metallicity, with implications for the formation of molecular gas and cold gas and for the star-formation history (see \citealt{Lebouteiller2017a}). The nature of such sources is still debated, though, and the modeling of their properties (including the luminosity in the soft X-rays which deposit their energy in the neutral gas), relies on the absorbing column density toward the X-ray source. An interesting prospect is thus to measure accurately the absorbing column density and ISM metallicity and ionization structure toward X-ray binaries (or toward OB stars in the same region) in nearby galaxies. The identification of compact objects in dwarf galaxies is important as such to probe potential intermediate-mass black holes and to understand whether they participate in the formation of supermassive black holes through coalescence (e.g., \citealt{Mezcua2019a}). Finally, another issue at stake is to understand the relative contribution of Wolf-Rayet stars vs.\ X-ray binaries in the nebular He\2\ emission in low-metallicity star-forming galaxies \citep{Schaerer2019a}.

\section{Gas flows and exchanges in the CGM}\label{sec:cgm}

\textbf{
\begin{itemize}
\item \textit{Is there enough CGM clouds to sustain star formation through accretion?}
\item \textit{What is the origin of high velocity clouds and how do they trade matter with the halo?}
\item \textit{How are halos energized?}
\end{itemize}}

\bigskip

No model of galaxy formation and evolution is complete without considering gas flows around galaxies. The CGM, in particular, stretches out to about the virial radii of galaxies and represents a key component of a galaxy’s matter budget that strongly influences its evolution over cosmic timescales. The evolution of galaxies is indeed thought to be regulated by a competition and balance between the gas accretion rate via cool gas streams infalling from the cosmic web, gas cooling in the halo, and mergers (the gas ``source term''), star formation in galaxies (the gas ``sink term''), and outflows driven by intense star formation and/or active galactic nuclei (the gas ``loss term''; e.g., \citealt{Bouche2010a,Richter2017a,Tumlinson2017a}). Of course, this combination of processes is too simplistic, there must be more complex microscopic processes that influence this cycle of accretion, star formation, and outflows. The question is then: what are those micro-physical processes that regulate the cooling and dissipation of the accreting and outflowing gas that maintain this apparent macroscopic gas balance in galaxies? The CGM in the halos of galaxies is a direct result of two of these processes -- gas accretion and outflows -- the two most important terms in the ``equation of gas balance in galaxies''. If we understand the nature and evolution of the CGM, we understand how galaxies acquire and lose their gas.

\subsection{The multi-phase CGM: a prime laboratory for galaxy evolution}

The CGM of galaxies is certainly not devoid of gas, perhaps containing up to approximately half of the total baryon content of the halo (e.g., \citealt{Werk2014a,Peek2015a}). The CGM is characterized by its multi-phase nature consisting of cold neutral/molecular gas clumps embedded in diffuse, highly-ionized gas filaments, with a wide range of temperature ($50-5\times10^6$\,K) and density ($10^{-5}-100$\cc) (e.g., \citealt{Jaffe2005a,Edge2010a,Salome2011a,Tremblay2012a,Emonts2018a}).

\begin{figure}
  \includegraphics[width=0.48\textwidth]{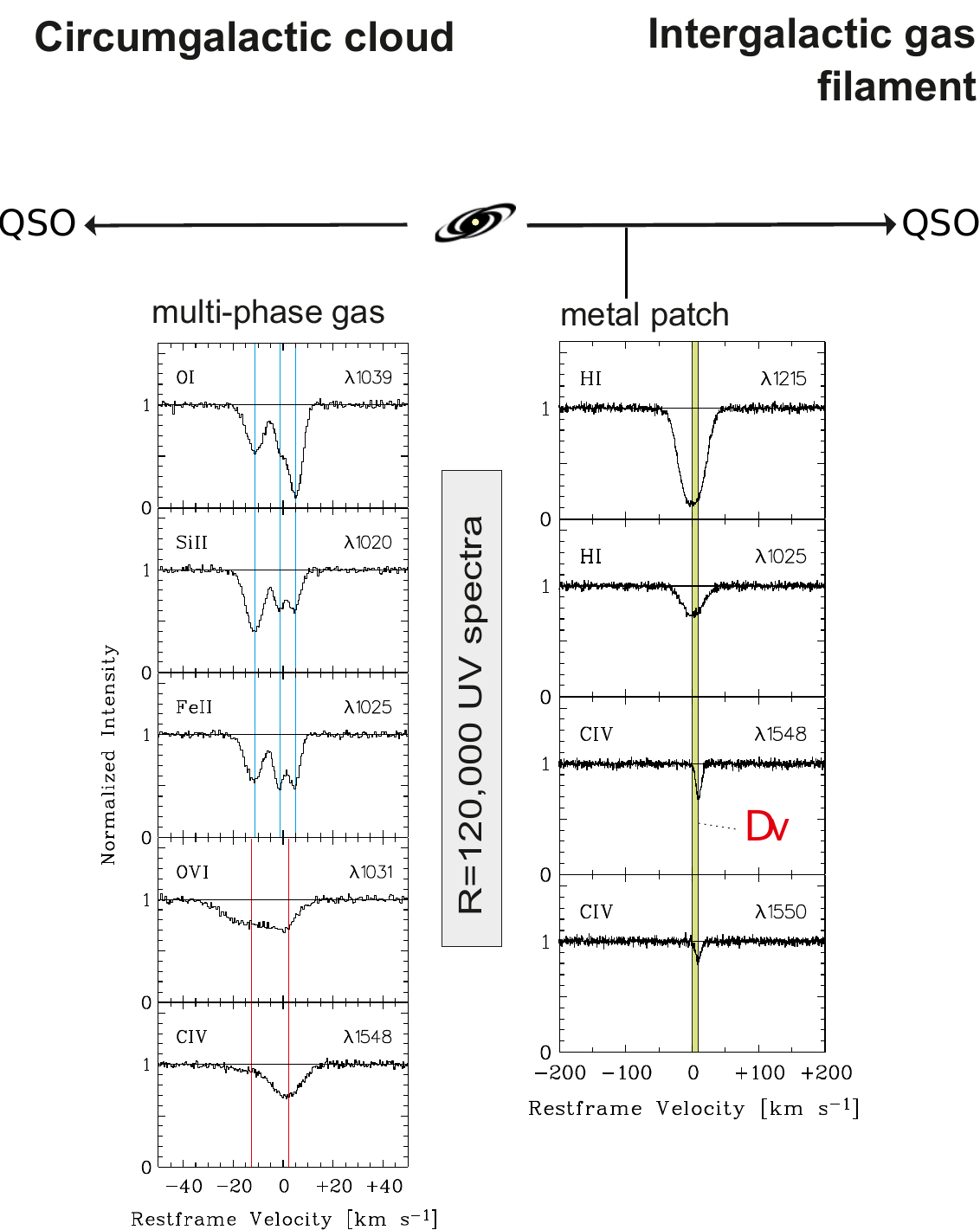}
  \caption{\small  Left-hand side: synthetic spectrum at $R=120\,000$ spectral resolution showing the typical velocity structure of a CGM absorber in several UV transitions. Due to the multi-phase nature of the absorber, the low ions O\1, Si\2, and Fe\2\ show velocity structure (blue vertical lines) that is different from that of the high ions O\6\ and C\4\ (red vertical lines). Right-hand side: velocity offset, $\Delta v$, (indicated by the green bar) between H\1\ and C\4\ in another synthetic ($R=120\,000$) indicating the kinematic displacement of a metal patch residing in an intergalactic filament as a result of an inhomogeneous metal mixing in the IGM. }\label{cgm}
\end{figure}

Absorption spectroscopy in the UV, where the quasars (or bright galaxies) are used as background sources to illuminate the foreground CGM of a galaxy (Fig.\,\ref{cgm}) represents the key method to study the physical nature of the CGM. 
The UV range covers in fact most of the diagnostic absorption lines to trace all of these gas phases from low to high redshifts with far-UV absorption but also emission lines (including far-UV rest-frame and redshifted extreme-UV transitions) down to very low gas column densities. Complementary methods include the detection of diffuse ionized gas emission, molecular gas emission, dust absorption, or hot gas in X-ray emission lines.

Past and present far-UV spectrographs (e.g., HST/STIS and HST/COS) have been used to study the CGM along quasar lines of sight, pioneering this observational approach to characterize the complex gas circulation processes around galaxies. The excellent sensitivity of COS enabled increasing the sample of absorbers at $z\lesssim1$ by about one order of magnitude as compared to previous studies \citep{Lehner2018a,Lehner2019a}. However, due to the limitations in sensitivity only a few quasar(s) per galaxy halo might be bright enough to spectroscopically investigate the CGM of foreground galaxies, hampering our knowledge of how the CGM functions as a dynamic reservoir for both infalling and outflowing gas. At present, only in M\,31 has it been possible to probe a significantly large ($\approx50$) number of QSOs, probing the CGM out to $1.5$ times the virial radius of the galaxy (\citealt{Lehner2015a,Howk2017a}, Lehner et al.\ in prep.). Increasing again the total number of absorbers by one or two orders of magnitude will be another game changer and will provide access to more reference sources such as M\,31. 

Another limitation has been the spectral resolution, which has not enabled us to kinematically disentangle the different gas phases, preventing accurate ionization models that are necessary to characterize the physical conditions in the gas and the role of this gas for galactic feedback. The conclusions that can be drawn from these simple experiments therefore remain afflicted with large uncertainties. As a result, our current understanding of the nature of the CGM is highly incomplete.

We now need to determine both the spatial distribution and large-scale kinematics of hydrogen and metal ions in the CGM around low- and high-redshift galaxies, as well as the physical conditions in the gas and its internal density structure. Because the enrichment of the IGM comes from galaxies that reside in the knots of the cosmological filaments, the metal distribution in the IGM might be inhomogeneous and patchy. Very high spectral resolution is required to systematically investigate velocity offsets between the H\1\ and metal ions (e.g., C\4, Si\3) that could indicate a poor metal mixing in the IGM.
Ionization conditions need to be determined for individual phases in order to provide reliable estimates of the total gas mass. Key diagnostic species range from molecular species such as H$_2$ to highly-ionized species such as O\6. At a spectral resolution of $\sim3$\,km\,s$^{-1}$ at $1000$\,\AA, profile fitting of absorption features from metal ions in the CGM directly delivers the temperature of absorbing gas (by resolving thermal line broadening), which then can be used together with the observed ion ratios to self-consistently model the ionization structure of CGM absorbers and their internal gas pressure. On the other hand, the analysis of fine-structure lines such as C\2$^{\star}$ helps to constrain the local cooling rates.

More than the distribution of metals and large-scale kinematics in the CGM, we also need to understand how galaxy halos are energized -- one of the crucial remaining frontiers in theoretical astrophysics. We are only starting to understand empirically how outflows from galaxies might be able to heat the halo gas. \cite{Hayes2016a} have, for instance, undertaken a comprehensive study of halos of star forming galaxies at $z\sim0.2$ with HST/ACS and COS. In this first study of a single star-forming galaxy, they have detected a halo of O\6\ $1032,1038$\,\AA\ emission. Since the O\6\ coronal doublet lines are the most significant coolant of gas at $T\sim10^{5-6}$\,K, this indicates that gas in this temperature range is cooling rapidly. The resolved O\6\ emission is extended over $26$\,kpc, i.e., ten times the size of the photoionized gas. About $1/6$ of the total O\6\ luminosity was estimated to come from resonantly scattered continuum radiation. From the spectra, \cite{Hayes2016a} derive a large column of O\6\ absorption which is outflowing at several hundred \kms. Since the UV spectrum of this galaxy resembles a stack of star-forming galaxies in the HST archive, this strongly suggests that this is a common phenomenology in star-forming galaxies. The mapping of this gas and its characterisation at high spectral resolution represents a crucial step in further constraining galaxy formation scenarios and provides the first direct evidence of how outflowing gas interacts with and heating the ambient halo gas (see also Sect.\,\ref{sec:layers}).

\subsection{The properties and role of high velocity clouds}

Sustainable star formation is achieved thanks to low-metallicity gas inflow. High velocity clouds (HVCs) are the prime candidates for such an inflow. HVCs are gaseous (apparently starless) clouds evolving in the halo of spiral galaxies and confined to the inner CGM. HVCs are, as such, useful probes of the gas exchanges between the disk and the halo (see \citealt{Lockman2019a}). While much progress has been done over the last decade or so notably concerning their distance and metallicity (e.g., \citealt{Lehner2011a,Lehner2012a,Wakker2007a,Wakker2008a,Fox2016a}), many questions remain open: (1) do HVCs provide enough mass to sustain star formation? (2) Can metallicity and abundance patterns be used to infer their origin and the chemical composition changes as it evolves through the halo? (3) How much dark matter do they contain? 

Several avenues can be considered to improve our knowledge on the properties and role of HVCs. We refer to \cite{Lockman2019a} for a review and we concentrate below on the need for UV observations. Since they can cover a broad range of ionization conditions, knowledge of the ionization corrections to be applied is key to determine the metallicities of HVCs. Since the gas may well not be in equilibrium a decisive way to free oneself from uncertainties is to observe all conceivable ionization states for a same atom, like, for instance [C\1, C\2, C\3, C\4], [Si\1, Si\2, Si\3, Si\4], or [N\1, N\2, N\3, N\5]. This requires the far-UV domain, especially access to N\2 at $1084$\,\AA\ and  O\6 at $1032$ and $1038$\,\AA. 

Determining the covering factor of HVCs as a function of distance and $z$-height by observing absorption lines toward a large sample of halo stars (B1 to B5, PAGB, BHB) with known distances (GAIA) would help infer the 3D structure of HVCs and determine whether the HVCs are disrupted and incorporated into the halo coronal gas as they fall or if they survive as neutral or ionized gas and reach the disk where they can fuel star formation. Only the H\1\ Lyman lines enable measuring $N({\rm H})$ down to $\lesssim10^{15}$\,cm$^{-2}$ (e.g., \citealt{Lehner2006a,Fox2006a,Zech2008a}).

\section{2020-2035 landscape}

The landscape of available observatories in the period 2020-2035 is not expected to change significantly as far as far-UV spectrographs or spectropolarimeters are concerned. Proposals for probe-class missions (e.g., CETUS for NASA, CASTOR for CSA) are being put forward to bridge the gap between HST and next generation UV missions but they should focus on surveys to complement those already planned in other wavelength ranges in the 2020s notably around the quest for dark energy. Other already proposed or planned mission accessing to the far-UV have requirements that do not enable the science objectives described in this paper (e.g., HabEx with $R<60\,000$ and $\lambda_{\rm min}=1150$\,\AA\ or WSO-UV with an effective area $A_{\rm eff}\lesssim10^3$\,cm$^{-2}$, $R<50\,000$, and $\lambda_{\rm min}=1150$\,\AA). There is currently to our knowledge no plan for a UV spectrolarimeter apart from the proposed NASA mission LUVOIR, which is the only planned mission with requirements strongly overlapping with the ones we advocate for (Sect.\,\ref{sec:req}).

This does not imply, however, that the science topics described in this document will not be tackled until 2035 and afterwards.
ALMA is expected to continue characterizing the molecular chemistry in the ISM, but is unable to access directly the reservoir of H$_2$ or to probe the various ISM phases. Future IR space missions may, on the other hand, provide access to many ISM tracers, including molecular and atomic fine-structure lines in a wide range of objects. Hence the ESA M5 proposed mission SPICA, if selected, or the NASA Origins mission submitted to the 2020 Decadal survey (with high spectral resolution $R>10^{5}$ across $100-200$\mic) could greatly improve the knowledge of the ISM in general, in particular of PDRs and molecular clouds in the Milky Way and in nearby galaxies. However, these missions will not provide access to many important transitions (e.g., cold H$_2$, H\1, hot gas) and are expected to primarily tackle the physics in star-forming regions rather than more diffuse gas, quiescent regions (i.e., irradiated by the general interstellar radiation field), or potential pristine gas pockets or HVCs. Polarimetry is proposed for both missions but only in large bands with no spectral resolution. 

Furthermore, MUSE on the VLT is enabling a significant leap for the characterization of the ionized gas in nearby objects and such instruments provide important comparisons with ISM properties derived in the far-UV, for instance regarding the ISM enrichment. The ELT will focus more on resolved stellar populations or star-forming systems and high-$z$ galaxies but will be an extremely valuable observatory to use in synergy with a dedicated far-UV mission, concerning the properties of the stellar population in various wavelength ranges but also the interplay between the ISM and star formation. SKA will be a game changer as far as the distribution and properties of atomic hydrogen in and around galaxies are concerned (small-scale structures, dynamics, matter exchange around galaxies), and a far-UV mission should be capable of characterizing the metals and molecular gas in low surface brightness regions and filaments that will be discovered with SKA.


In summary, the landscape of existing, planned, or proposed missions until and after 2035 will enable a multi-wavelength view of galaxies, but missing the far-UV domain and its unique tracers. There exists, therefore, the opportunity that a far-UV mission may fill the gap in this landscape, with great potential synergies.

\section{Conclusion and requirements}\label{sec:req}

\begin{table*}
  \caption{High-level requirements. }\label{tab:req}
  \begin{tabular}{l|p{10cm}}
    \hline
    \hline
    {\bf Requirement} & {\bf Justification and comments}\\
    \hline
    \textit{\textbf{Wavelength range}} & \\
    \hline   
 $\lambda_{\rm min} {\rm (strict)}=1020$\,\AA & H\1\ Lyman $\beta$ $1025$\,\AA; strongest \hmol\ Lyman bands: $1030-1155$\,\AA; CO up to  $1455$\,\AA; O\6 $1032$\,\AA; N\2 $1084$\,\AA; Ar\1 $1048, 1066$\,\AA; O\1 $1039, 1026$\,\AA. \\
 $\lambda_{\rm min} {\rm (preferred)} = 900$\,\AA & H\1\ Lyman series down to $912$\,\AA; \hmol\ lines $912-1155$\,\AA; CO lines $912-1455$\,\AA; many O\1\ lines of various strengths $916-988$\,\AA. Rest-frame Lyman continumm (no resolution requirement). \\
    $\lambda_{\rm max} = 3100$\,\AA & Mg\2 $2800$\,\AA; Mg\1\ $2853$\,\AA; OH lines $3072-3082$\,\AA. \\
    \hline
    \textit{\textbf{Resolution}} & \\
    \hline 
 $R=\lambda/\Delta\lambda {\rm (strict)} >120\,000$ & Resolve line profiles from cold \hmol\ with $T\simeq100$\,K and $v_{\rm turb}\simeq1.2$\,\kms; resolve line profiles, separate thermal and turbulent contributions for warm gas: Fe\2\ with $T\simeq6500$\,K and $v_{\rm turb}\simeq1.0$\,\kms; separate velocity components with  $\Delta{v}\simeq3$\,\kms; resolve profiles from rotational levels in \hmol\ and CO bands; resolve isotopical shifts of atomic species.   \\ 
    $R {\rm (preferred)} > 200\,000$ & Resolve line profiles from cold gas with $v_{\rm turb}\leq1.0$\,\kms; resolve isotopes with $\Delta{v}\leq1.5$\,\kms. \\
    $R\sim30\,000$ & FUSE/COS-quality spectra toward individual stars a few Mpc away; minimum resolution for GSA for absorption lines. \\
    Spatial (PSF or aperture size) $=10$\,mas  & Observe individual lines of sight toward stars in galaxies few Mpc away; tomographic mapping of chemical properties.  \\
    \hline
    \textit{\textbf{Sensitivity}} & \\
    \hline   
SNR $> 500$ & Detect faint features in a reasonable amount of time (signatures of GSA in Milky Way, scarce elements, fine-structure lines in extragalactic ISM...). This implies achieving detectors with limited fixed-pattern noise and that can deal  with high count rates.   \\
    $A_{\rm eff}>6\,000$\,cm$^{2}$ & $3\times$ the HST/COS value. Needed to reach $A_V>4$. Achieved with a telescope $5$\,m (resp. $8$\,m)  if the efficiency (optical $\times$ detector) reaches $3$\% (resp. $1.3$\%).  \\
    $A_{\rm eff}>40\,000$\,cm$^{2}$ & $R\sim10^5$ spectra with main features toward galaxies $1-3$\,Mpc away; current LUVOIR-B designs for LUMOS and POLLUX;   \\
\hline
    magnitude [AB] $\approx27$  & Individual stars in I\,Zw\,18 ($18$\,Mpc) with $R\approx30\,000$. \\
                                  \hline
   \textit{ \textbf{Observational modes}} & \\
    \hline
    Apertures: MOS \& pinhole/long-slit & Ability to observe single stars / QSOs either sequentially or simultaneously; small enough to warrant required high spectral resolution. \\
Polarimetric mode & Magnetic fields. Linear (QU) polarization required. Circular(V)+linear prefered for evidence of chirality. Ability to use full spectroscopic mode. \\
    \hline
  \end{tabular}
\end{table*}

The science requirements described in the paper dictate the instrumental specifications in Table\,\ref{tab:req}. Concerning the polarimetric capability, since birefringent materials do not transmit light below $1200$\,\AA, new techniques have to be developped. The ability to operate in pure spectroscopic mode is important. Considering the sensitivity and SNR requirements, detectors have to be able to withstand high count rates without saturating. 

Long, narrow-slit spectroscopy or small aperture pinholes are well adapted to the observation of single stars (in the Milky Way or in nearby galaxies) or QSO lines of sight. A multi-object or integral field spectrograph would be required for crowded fields or extended regions, in particular for nearby galaxies (either multiple stars within galaxy or multiple QSO sightlines intersecting ISM+CGM). 

All in all, the proposed mission corresponds to an L-class asssuming current and proposed technology developments such as those presented in the NASA Large Ultraviolet/Optical/Infrared Surveyor (LUVOIR) report \citep{TheLUVOIRTeam2018a}. LUVOIR is one of four missions of the flagship class that are being studied by NASA in the framework of the US 2020 Decadal Survey. LUVOIR is a multi-purpose observatory, designed to address very ambitious scientific questions at the core of modern Astrophysics, astrophysics, thanks to four instruments. The European community is involved in the LUVOIR project, through the proposal of one of the four instruments: POLLUX (PIs J.-C.\ Bouret \& C.\ Neiner), a high-resolution spectropolarimeter operating at UV wavelengths, designed for the $15$-meter primary mirror option (LUVOIR-A). POLLUX, whose current development is funded by CNES is at present the only non-US instrument proposed for LUVOIR. POLLUX uses multiple reflections to circumvent the transmission issue, and is well adapted to the requirements needed for the magnetic field science case in this document \citep{Muslimov2018a,LeGal2019a}. Technology challenges will include characterization of reflective coating materials (see \citealt{Muslimov2018a}). 

Globally, the POLLUX and LUMOS instruments proposed for LUVOIR enable the science objectives described in this paper. We therefore advocate for an ESA contribution to an international effort such as LUVOIR around the POLLUX instrument. It should be emphasized, however, that since proper coatings have to be used in order to obtain useful effective areas around and below $1000$\,\AA, LUVOIR is itself not completely optimized for this specific wavelength range and the question of having a dedicated far-UV mission should be raised.

\setlength{\bibsep}{0.0pt} 


\end{document}